\documentclass[lettersize,journal]{IEEEtran}

\IEEEoverridecommandlockouts

\usepackage[cmex10]{amsmath}
\usepackage{amsthm}
\usepackage{amsmath}
\usepackage{algorithmic}
\usepackage{array}
\usepackage{bm,cite}
\usepackage{amssymb}
\usepackage[tight,footnotesize]{subfigure}
\usepackage{url,caption, color, enumerate, epsfig}
\usepackage{algorithm}

\usepackage{epsfig,rotating,setspace,latexsym,amsmath,epsf,amssymb,amsfonts,bm,theorem,cite,authblk, bbm,color, hyperref, multirow, caption}

\IEEEoverridecommandlockouts
\allowdisplaybreaks
\title{Jamming Attacks on Decentralized Federated Learning in General Multi-Hop Wireless Networks}
\begin{document}
\author[1]{Yi Shi}
\author[2]{Yalin E. Sagduyu}
\author[2]{Tugba Erpek}

\affil[1]{\normalsize  Virginia Tech, Commonwealth Cyber Initiative, Arlington, VA, USA}
\affil[2]{\normalsize  Virginia Tech, National Security Institute, Arlington, VA, USA}
\maketitle



\begin{abstract}
Decentralized federated learning (DFL) is an effective approach to train a deep learning model at multiple nodes over a multi-hop network, without the need of a server having direct connections to all nodes. In general, as long as nodes are connected potentially via multiple hops, the DFL process will eventually allow each node to experience the effects of models from all other nodes via either direct connections or multi-hop paths, and thus is able to train a high-fidelity model at each node. We consider an effective attack that uses jammers to prevent the model exchanges between nodes. There are two attack scenarios. First, the adversary can attack any link under a certain budget. Once attacked, two end nodes of a link cannot exchange their models. Secondly, some jammers with limited jamming ranges are deployed in the network and a jammer can only jam nodes within its jamming range. Once a directional link is attacked, the receiver node cannot receive the model from the transmitter node. We design algorithms to select links to be attacked for both scenarios. For the second scenario, we also design algorithms to deploy jammers at optimal locations so that they can attack critical nodes and achieve the highest impact on the DFL process. We evaluate these algorithms by using wireless signal classification over a large network area as the use case and identify how these attack mechanisms exploits various learning, connectivity, and sensing aspects. We show that the DFL performance can be significantly reduced by jamming attacks launched in a wireless network and characterize the attack surface as a vulnerability study before the safe deployment of DFL over wireless networks.
\end{abstract}

\begin{IEEEkeywords}
Decentralized federated learning, jamming attack, wireless network, network security 
\end{IEEEkeywords}

\section{Introduction}
A machine learning model is typically trained at one device by using the data collected or made available at this device while some applications may leverage data collected at multiple devices. For such applications, a server can be employed to collect data and to train a machine learning model. However, such an approach may have high communication overhead due to large amount of raw data and may raise privacy concerns on sharing the local data from individual devices. To address these concerns, \emph{federated learning} (FL) \cite{McMahan17:FL} has been formulated to train a machine learning model jointly among multiple clients. In FL, each client does not send its local data to a server. Instead, it trains and sends its local model to the server. The server performs some form of federated averaging algorithm to obtain a global model and sends this model to all clients for the next round of training. The final global model can achieve similar performance as the model trained by using all clients' data at one device. FL has the advantage of data privacy, low communication overhead, and low computational complexity at each node (due to splitting the training task to multiple clients) \cite{Survey0:FL, Survey1:FL, Survey2:FL, Survey3:FL}.



\begin{figure}
\centering   \includegraphics[width=\columnwidth]{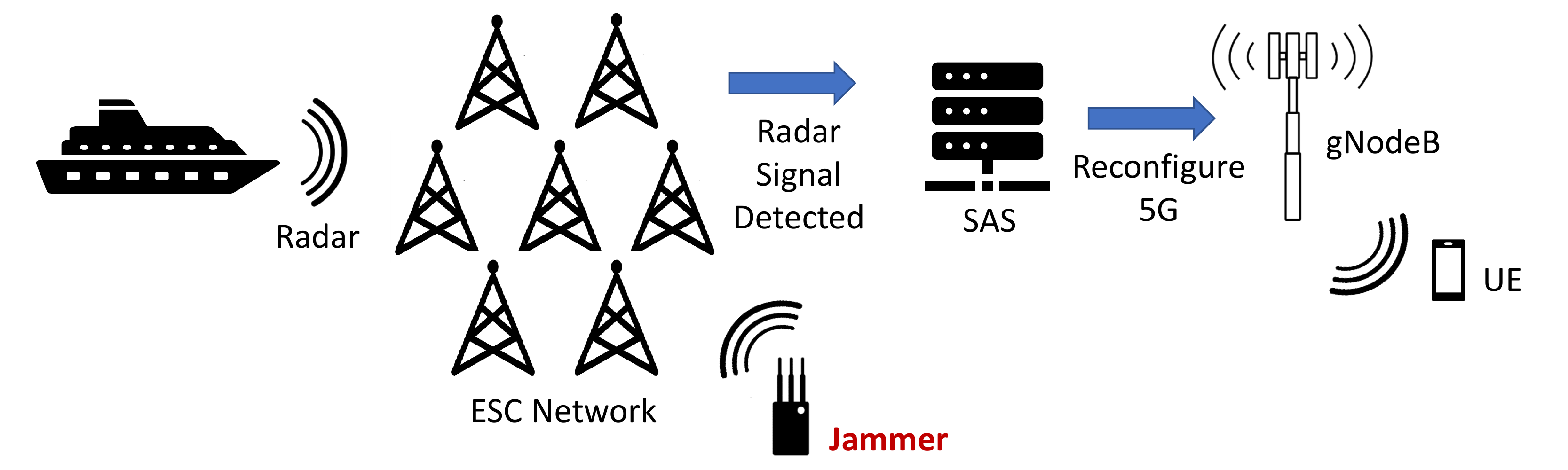}
   \caption{The wireless sensor network for ESC operating in the CBRS band under jamming attacks.}
\label{fig:CBRS}
\end{figure} 

In this paper, we consider a \emph{distributed spectrum sensing} problem as the example task for machine learning (see Fig.~\ref{fig:CBRS}). One real-world use case is the Environment Sensing Capability (ESC) for 3.5GHz Citizens Broadband Radio Service (CBRS) \cite{FCC}, where the incumbent user (namely, radar) and the NextG communication system need to share the spectrum band in the same region. To that end, the ESC sensors need to detect the incumbent signal and inform the Spectrum Access System (SAS) to configure the 5G communications system to prevent interference with the incumbent user \cite{DL:CBRS}. 
Instead of building a deep learning model, namely a deep neural network (DNN), that works for a particular ESC node, it is beneficial to build a general model that can work for any ESC node. For this purpose, multiple nodes collect (spectrum sensing) data at the different locations and jointly train a DNN. FL may not be applicable to this problem setting since (i) there is no server in the network and (ii) we may not let a node to work as a server since a server should be able to communicate to clients directly. One approach for this problem setting is \emph{decentralized federated learning} (DFL) \cite{Shi22:DFL, Liu22:DFL, Lalitha:DFL}, where each node works as a server for its neighbors and performs federated averaging for its local model and models from its neighbors. Assuming the entire network is connected (potentially over multiple hops), each node can eventually have information for any other node via either direct connections or multi-hop paths.

While performing complex tasks, machine learning is known to be vulnerable to attacks that manipulate its training data or test data. Attacks based on \emph{adversarial machine learning} have been extensively studied for  wireless systems  \cite{Adesina:AML, BookchapterAML}.
FL is also susceptible to a variety of attacks \cite{MEC:FL, Freerider, Mittal, Lyu20:FL-attack, FLSecurity}. One particular attack that is unique for wireless systems is the \emph{jamming attack} that has been studied in terms of disrupting the model exchange between the server and the clients of FL \cite{Shi22:FL-Attack, Shi22:FL-Attack2}.
On the other hand, the attack surface for DFL is not well understood  yet. In this paper, we consider an adversary that can jam some links in the wireless network. Once jammed, nodes cannot exchange their local models over these links. We consider two scenarios. In the first scenario, an adversary can jam any link in the network subject to a jamming budget. The problem is how to select the jammed links to maximize the impact on the DFL process. In the second scenario, a jammer has a limited jamming range (i.e., it can only jam receivers within some distance). Thus, we need to first deploy jammers in critical locations and then jam the links that fall within the jamming range. 
The contributions of this paper are summarized as follows.
\begin{itemize}
\item We design jamming attack algorithms for Scenario~$1$, where two end nodes of a link under attack cannot exchange their local models. The entire network may be broken as several connected components and a node in one connected component cannot collect information from other connected components. Thus, we design the minimum cut based attack (MCBA) to maximize the impact of jamming attack.  To reduce the complexity, we also design the node degree based attack (NDBA).

\item We design jamming attack algorithms for Scenario~$2$, where a set of jammers are deployed in the network. For a directional link $n_1 \to n_2$, if $n_2$ is within the jamming range of a jammer, this jammer can attack this link and node $n_2$ cannot receive any model from node $n_1$.
We design the two attacks, MCBA and NDBA, also for Scenario~2. Before links are selected, jammers are deployed to cover critical nodes so that links ending at these nodes can be attacked.

\item We compare the performance of jamming attack algorithms for both scenarios. First, we show that DFL can train a DNN with high accuracy in a wireless network. Then, we show that both attack algorithms are very effective in terms of reducing the DNN accuracy and increasing the learning time while a random attack algorithm cannot achieve this effect. Since the performance of the NDBA is similar to that of the MCBA, it is viable to adopt the NDBA to launch low-complexity attacks.
\end{itemize}

The rest of the paper is organized as follows. 
Section~\ref{sec:problem} describes the signal classification problem and the DFL approach. 
Section~\ref{sec:attack} presents the attack scenarios and attack algorithms for DFL. 
Section~\ref{sec:result} evaluates the attack performance. 
Section~\ref{sec:conc} concludes this paper.

\section{DFL for Signal Classification}
\label{sec:problem}

We consider a \emph{wireless signal classification} problem, where a transmitter sends data via either BPSK or QPSK and the received signals are classified to labels `BPSK' and `QPSK'. For wireless networks with complex channel characteristics, it has been a common practice to use \emph{deep learning }to improve the accuracy of wireless signal classification \cite{DL:wireless1, DL:wireless2}.
A DNN can be trained for signal classification using phase shift and power level in the received data as the features.
To avoid the bias due to the particular channel condition experienced when collecting data at one location, multiple nodes can sense the spectrum, collect data, and work together to train a model jointly.

\begin{figure}
\centering  		
\includegraphics[width=0.7\columnwidth]{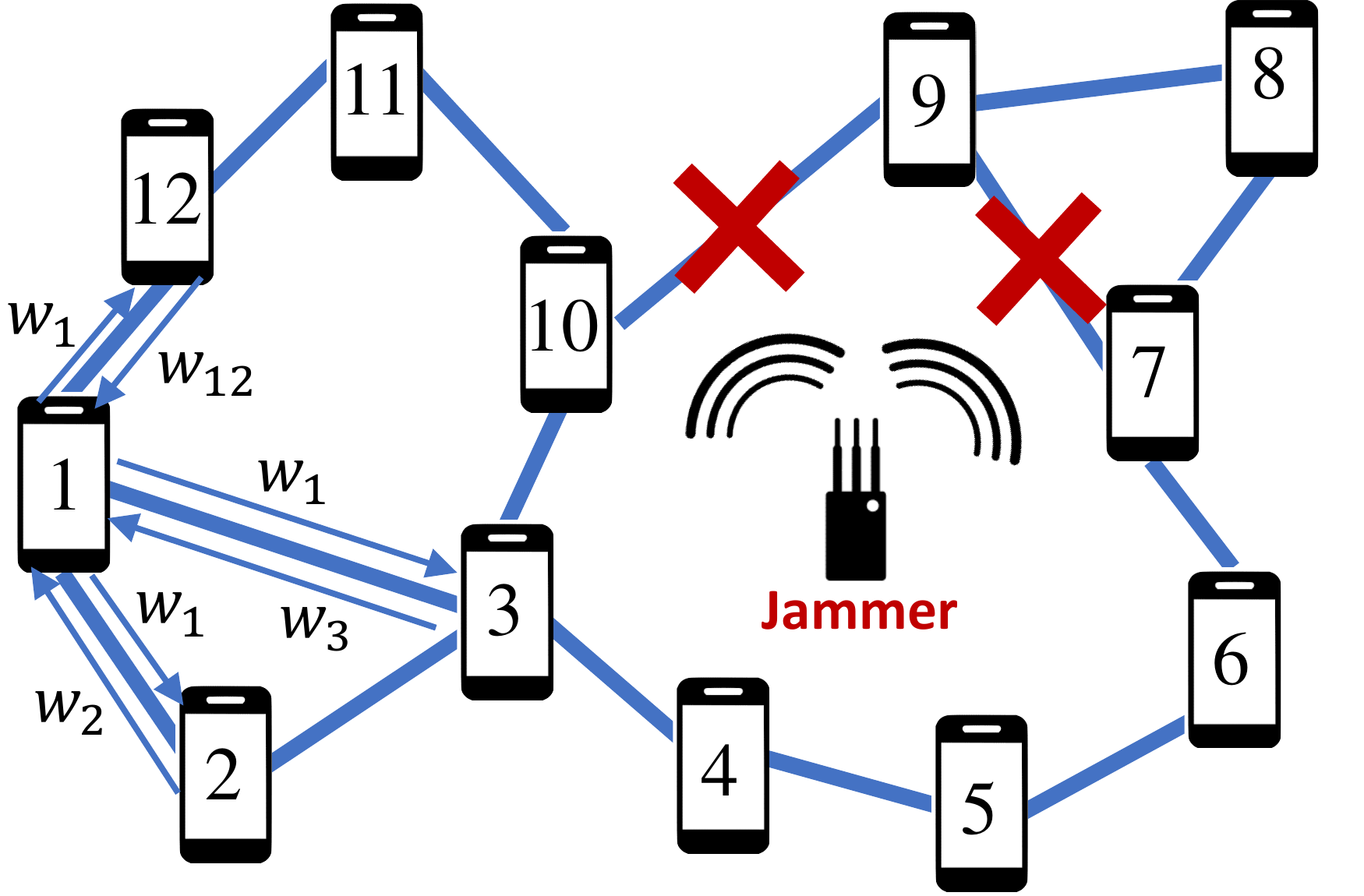}
\caption{DFL in a multi-hop network (e.g., node~$1$ broadcasts its model $w_1$ to its neighbor nodes $2, 3$, and $12$ and receives their models $w_2$, $w_3$, and $w_{12}$ in a multi-hop network) in the presence of jammer (e.g., jams links $7 \to 9$ and $9 \to 10$.}
\label{fig:general}
\end{figure}

FL allows multiple clients to train a model together without sharing their data. In the typical setting, FL relies on a server with direct connections to all nodes to obtain a global model by federated averaging. However, we do not assume server. The reason is that we may not specify a node as the server since there may not be a node with direct connections to all other nodes due to limited communication range (see Fig.~\ref{fig:general} for an example network topology). Also, it is better to avoid a dedicated server as a single point of failure.
Therefore, we consider DFL to train a model in a multi-hop wireless network environment without requiring a server.
DFL follows the following steps.
\begin{itemize}
\item Each node starts with a random DNN model at the beginning.
\item During each round, each node performs the following.
\begin{itemize}
\item Each node trains (updates) its model using its local data.
\item Each node sends its model to its neighbors and receives models from its neighbors.
\item Each node performs federated averaging on its model and its neighbors' models to obtain an updated model, which is used as the initial model for the next round.
\end{itemize}
\end{itemize}
The DFL process ends if the improvement at all nodes is limited or if the maximum number of rounds is reached. To train a good model by DFL, we assume that the entire network is connected so that each node can eventually collect information from all other nodes via either direct connections or multi-hop paths. 

\section{Jamming Attack on DFL}
\label{sec:attack}

An adversary can launch a \emph{jamming attack} to break communication links. As a consequence, a node cannot receive a neighbor's model transmitted over a jammed link. If the entire network is still connected, this node can collect information in these missing models by alternative paths but the learning process will be longer. Otherwise (namely, if the network is divided into multiple connected components), a node can only collect information from nodes in the same connected component and lose the information from nodes in other connected components. Then, the final model may be much worse due to the missing information.

\begin{algorithm}[t]
    \caption{MCBA algorithm for Scenario~$1$.}
    \label{alg:MC}
    \begin{algorithmic}[1]
        \STATE Calculate the minimum cut for the bidirectional connectivity graph of nodes.

        \IF{the budget $M_b$ is less than the number of links $N_{MC}$ in the minimum cut} 
            \STATE For each link in the minimum cut, denote $(d_1, d_2)$ as the degrees of its two end nodes, where $d_1 \le d_2$.

            \STATE Among these links, select $M_b$ links with small $d_1$ values. To break the tie among some links with the same $d_1$, select links with small $d_2$ values.
        \ELSE 
            \STATE Select all links in the minimum cut.
            
            \STATE For each link not in the minimum cut, denote $(d_1, d_2)$ as the degrees of its two end nodes, where $d_1 \le d_2$.

            \STATE Among these links, select $M_b - N_{MC}$ links with small $d_1$ values. To break the tie among some links with the same $d_1$, select links with small $d_2$ values.
        \ENDIF
    \end{algorithmic}
\end{algorithm}

\begin{algorithm}[t]
    \caption{NDBA algorithm for Scenario~$1$.}
    \label{alg:degree}
    \begin{algorithmic}[1]
        \STATE For each link, denote $(d_1, d_2)$ as the degrees of its two end nodes, where $d_1 \le d_2$.

        \STATE Select $M_b$ links with small $d_1$ values. To break the tie among some links with the same $d_1$, select links with small $d_2$ values.
    \end{algorithmic}
\end{algorithm}

There are \emph{two attack scenarios} that we consider in this paper. For the first scenario, an adversary can jam any links in the network. Once jammed, the two end nodes of any jammed link cannot exchange their models. In this scenario, the connections in the network can be modeled as a bidirectional graph.
For the second scenario, a jammer has a limited jamming range $R_J$ and can only jam nodes within its jamming range. Once a link is jammed, the receiver node cannot receive the transmitter node's model. In this scenario, the connections in the network can be modeled as a directional graph. For both scenarios, there is a \emph{jamming budget}, i.e., there is an upper bound on the number of jammed links. We denote this budget as $M_b$ for the first scenario and $M_d$ for the second scenario.

We now present the jamming attack algorithms for the first scenario.
To maximize the impact of jamming, in terms of longer learning time and worse accuracy of final models, the jammed links should be carefully selected.
As we pointed out earlier, it would be more effective if the entire network is broken as multiple connected components under this attack. We can design an attack based on the \emph{minimum cut} concept, namely the MCBA. A cut is a set of links that breaks a network into multiple connected components. A minimum cut is a cut with the minimum number of links. That is, an adversary first jams links in a minimum cut and then jams other links.
For links in the same category, the adversary jams links with low node degrees to maximize the impact. The details of the MCBA are shown in Algorithm~\ref{alg:MC}.

\begin{algorithm}[t]
    \caption{Minimum cut based deployment algorithm.}
    \label{alg:deploy-MC}
    \begin{algorithmic}[1]
        \STATE Calculate the minimum cut for the directional connectivity graph of nodes.

        \WHILE{the number of deployed jammers is less than $N_A$}
            \STATE Find the node $n_1$ with the minimum in-degree among nodes in the minimum cut and not covered by deployed jammers. If not found, break. 

            \STATE Find the node with the next minimum in-degree among nodes in the minimum cut, not covered by deployed jammers, and with a distance to $n_1$ no more than $2 R_J$. If not found, find the node with the minimum in-degree among nodes not in the minimum cut, not covered by deployed jammers, and with a distance to $n_1$ no more than $2 R_J$. Denote this node as $n_2$.

            \IF{$n_2$ is not found}
                \STATE Deploy a jammer at the location of $n_1$.
            \ELSE
                \STATE Deploy a jammer at the midpoint of $n_1$ and $n_2$.
            \ENDIF
        \ENDWHILE

        \WHILE{the number of deployed jammers is less than $N_A$}
            \STATE Find the node with the minimum in-degree among nodes not covered by deployed jammers. If not found, break. Otherwise, denote this node as $n_1$.

            \STATE Find the node $n_2$ with the next minimum in-degree among nodes not covered by deployed jammers and with a distance to $n_1$ no more than $2 R_J$. 

            \IF{$n_2$ is not found}
                \STATE Deploy a jammer at the location of $n_1$.
            \ELSE
                \STATE Deploy a jammer at the mid point of $n_1$ and $n_2$.
            \ENDIF
        \ENDWHILE
    \end{algorithmic}
\end{algorithm}

The complexity of calculating a minimum cut is high for a large network. To reduce complexity, the adversary may consider the NDBA, where the \emph{node degree} is used to select links for the jamming attack. That is, the adversary jams links with low node degrees to maximize the impact of this attack. The details of the NDBA are shown in Algorithm~\ref{alg:degree}.

\begin{algorithm}[t]
    \caption{MCBA algorithm for Scenario~$2$.}
    \label{alg:MC2}
    \begin{algorithmic}[1]
        \STATE For each link $n_1 \to n_2$ in the minimum cut, where $n_2$ is covered by some jammer(s), denote $d_2$ as the in-degree of node $n_2$.

        \STATE Among these links, select up to $M_d$ links with small $d_2$ values. Denote the number of selected links as $M_1$.
            
        \IF{$M_d > M_1$}
            \STATE For each link $n_1 \to n_2$ not in the minimum cut, where $n_2$ is covered by some jammer(s), denote $d_2$ as the in-degree of node $n_2$.

            \STATE Among these links, select up to $M_d - M_1$ links with small $d_2$ values. 
        \ENDIF
    \end{algorithmic}
\end{algorithm}


\begin{algorithm}[t]
    \caption{Node degree based deployment algorithm.}
    \label{alg:deploy-degree}
    \begin{algorithmic}[1]
        \WHILE{the number of deployed jammers is less than $N_A$}
            \STATE Find the node with the minimum in-degree among nodes not covered by deployed jammers. If not found, break. Otherwise, denote this node as $n_1$.

            \STATE Find the node with the next minimum in-degree among nodes not covered by deployed jammers and with a distance to $n_1$ no more than $2 R_J$. Denote this node as $n_2$.

            \IF{$n_2$ is not found}
                \STATE Deploy a jammer at the location of $n_1$.
            \ELSE
                \STATE Deploy a jammer at the midpoint of $n_1$ and $n_2$.
            \ENDIF
        \ENDWHILE
    \end{algorithmic}
\end{algorithm}

\begin{algorithm}[t]
    \caption{NDBA algorithm for Scenario~$2$.}
    \label{alg:degree2}
    \begin{algorithmic}[1]
        \STATE For each link $n_1 \to n_2$, where $n_2$ is covered by some jammer(s), denote $d_2$ as the in-degree of node $n_2$.

        \STATE Select up to $M_d$ links with small $d_2$ values. 
    \end{algorithmic}
\end{algorithm}

For the second scenario, there are two steps for launching attacks. In the first step, we need to deploy $N_A$ jammers at critical locations so that in the second step, these jammers can jam up to $M_d$ critical links to maximize the impact of attacks. We can still design the jamming attack algorithm based on the minimum cut concept. In the first step, we aim to deploy jammers to first cover nodes appeared in the minimum cut (Category 1), and then cover the remaining nodes (Category 2), where a node is covered by a jammer if it is within the jamming range of this jammer. For nodes in the same category, we aim to deploy jammers to cover nodes with low degrees.
The details are shown in Algorithm~\ref{alg:deploy-MC}. A jammer's location may be determined as the location of a node by Algorithm~\ref{alg:deploy-MC} (see line~$6$). If a jammer cannot be placed at the same location of a node, it should be deployed close to that location. Once jammers are deployed, they can jam up to $M_d$ links selected by Algorithm~\ref{alg:MC2}.

We can again reduce the complexity 
by considering the node degree only to deploy jammers  and to select links to be jammed.
The details are shown in Algorithms~\ref{alg:deploy-degree} and \ref{alg:degree2}.

\section{Performance evaluation}
\label{sec:result}

To evaluate the effect of jamming attacks on DFL, we consider a transmitter that can transmit using either QPSK or BPSK modulation. We analyze phase shift and power level for raw I/Q data to obtain features that are used as the input to the DNN for DFL. We consider features for $16$ bits as one sample. For BPSK, there are $32$ features. For QPSK, we calculate two sets of phase shift and power level for $2$ bits so that the number of features in a sample is also $32$. The output of the DNN is the classification label, i.e., `BPSK' or `QPSK'. We use feedforward neural network (FNN) as the DNN with parameters shown in Table~\ref{table:FFN}.

\begin{table}
\caption{FNN properties.}
\centering
{\small
    \begin{tabular}{c|c}
        Input size & 32 \\ \hline
	Output layer size & 2 \\ \hline
        Hidden layer sizes & $128, 64, 32$ \\ \hline
        Dropout rate & $0.2$ \\ \hline
        Activation function & ReLU (hidden layer)\\ & Softmax (output layer) \\ \hline
        Loss function & Crossentropy \\ \hline
        Optimizer & RMSprop \\ \hline
        Number of parameters & 14,626
    \end{tabular}
}
\label{table:FFN}
\end{table}

\begin{figure}
  \centering
  \includegraphics[width=1\columnwidth]{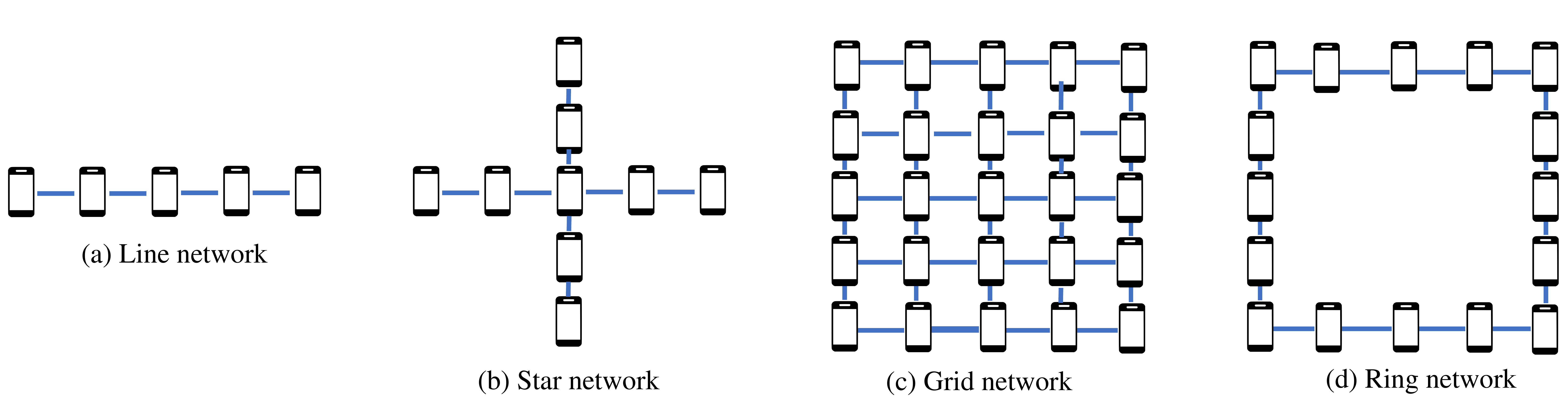}
   \caption{Line, star, grid, and ring network topologies.}
\label{fig:topologies} 
\end{figure} 

The transmitter is deployed at $(0, 0)$ while receiver nodes (namely, network clients) are deployed in $[200, 1000] \times [200, 1000]$. The communication range of these nodes is $200$. We consider different network topologies for nodes, including line, star, grid, and ring networks shown in Fig.~\ref{fig:topologies}, as well as a random network. When we consider a limited jamming range in Scenario~$2$, we set this range as $200$ and the number of jammers is 2. The number of attacked links is set approximately $25\%$ of all links. Note that we consider bidirectional links for Scenario~$1$ and directional links for Scenario~$2$. To compare the effect of these two scenarios, we set $M_d = 2 M_b$. The detailed setting for the networks is provided as follows.
\begin{itemize}
\item In the \emph{randomly generated network}, locations are randomly selected for $20$ nodes. For the maximum number of attacked links, we set $M_b = 10$ and $M_d = 20$.

\item In the \emph{line topology}, except two end nodes, all other nodes are connected with two neighbors. We assume that the line topology has five nodes at $(200i, 600)$ for $i=1, 2, \cdots, 5$. For the maximum number of attacked links, we set $M_b = 2$ and $M_d = 4$.

\item The \emph{star topology} has two lines crossing at the central node. We assume that the star topology has nine nodes at $(200i, 600)$ for $i=1, 2, \cdots, 5$ and $(600, 200i)$ for $i=1, 2, 4, 5$. For the maximum number of attacked links, we set $M_b = 2$ and $M_d = 4$.

\item In the \emph{grid topology}, each node is connected with up to four neighbors at left, right, up and down. In particular, we assume that the grid topology has $25$ nodes at $(200i, 200j)$ for $i, j \in {1, 2, \cdots, 5}$. For the maximum number of attacked links, we set $M_b = 10$ and $M_d = 20$.

\item In the \emph{ring topology}, all nodes are connected with two neighbors. We assume that the ring topology has $16$ nodes by removing $9$ nodes in the center from the grid topology. For the maximum number of attacked links, we set $M_b = 4$ and $M_d = 8$.
\end{itemize}

Each node collects $1000$ samples as its local data. To measure the performance, the test data is built by including $100$ samples from each node. We use this test data to ensure that the trained models can achieve a good performance not only for its own data but also for the data of other nodes. 
Although such a test dataset may not be available in practice, e.g., nodes do not want to share data due to privacy, we can still use it for the purpose of evaluating the performance of DFL under attack. To compare networks with different numbers of nodes in different topologies, we calculate the average accuracy over all nodes. We observe that this average accuracy fluctuates over rounds. Thus, we analyze when the best average accuracy converges. In particular, the best average accuracy in round $t$ is the maximum value of average accuracy measured in rounds $1$ to $t$. We claim convergence if the best average accuracy does not increase by a small value $\varepsilon$ for continuous $m$ rounds. We set $\varepsilon = 0.01$ and $m = 100$ for numerical results. We also study the minimum accuracy among all nodes. In addition to the MCBA and the NDBA, we also consider the following benchmark schemes.
\begin{itemize}
\item \emph{No attack}. All links are available for DFL.

\item \emph{Attack all links}. Each node can only train a model based on its own data.

\item \emph{Random attack}. Links are randomly selected for attack.
\end{itemize}

\begin{table}
	\caption{Attack performance for a random topology.}
	\centering
	{\vspace{0.2cm}
	\footnotesize
		\begin{tabular}{c|c|c|c}
Attack & Convergence & Minimum & Average \\
algorithm & time & accuracy & accuracy \\
\hline \hline
No attack & 497 & 94.00 & 94.94 \\ \hline
Attack all links & $\ge 1000$ & 50.00 & 71.95 \\ \hline
\multicolumn{4}{c}{Unlimited jamming range} \\ \hline
MCBA & 571 & 54.70 & 89.68 \\ \hline
NDBA & 571 & 54.70 & 89.68 \\ \hline
Random & 641 & 54.70 & 94.76 \\ \hline
\multicolumn{4}{c}{Limited jamming range} \\ \hline
MCBA & $\ge 1000$ & 51.10 & 76.77 \\ \hline
NDBA & $\ge 1000$ & 50.60 & 76.44 \\ \hline
Random & 842 & 87.50 & 93.04 
		\end{tabular}
	}
	\label{table:random}
\end{table}

Table~\ref{table:random} shows the results for a randomly generated network.
When there is no attack, DFL can converge at time $497$ (i.e., after $497$ rounds) with minimum accuracy $94.00\%$. 
If an adversary can jam all links, each node can only train a model using its own data. These models cannot converge in $1000$ rounds. The performance can be very bad for some node(s). The minimum accuracy is $50.00\%$, which is the same as the performance of random guessing. The average accuracy is $71.95\%$, which is not high. This result shows the need of DFL to train a good model jointly. 
For Scenario~$1$, all attacks increase the convergence time and reduce the performance.
The MCBA and the NDBA have the same performance and we find that they attack the same set of links.
The random attack, although is effective to reduce the minimum accuracy, is not very effective to reduce other nodes' accuracy. That is, the average accuracy is $94.76\%$, which is almost the same as the average accuracy of no attack $94.94\%$.
For Scenario~$2$, all attacks again reduce the performance.
The NDBA is slightly better than the MCBA. 
The random attack is not effective to reduce either the minimum accuracy or the average accuracy. 
Comparing two scenarios, the attacks for Scenario~$2$ are more effective because for this scenario, jammers do not need to always attack a link in both directions. 

\begin{table}
	\caption{Attack performance for the line topology.}
	\centering
	{\vspace{0.2cm}
	\footnotesize
		\begin{tabular}{c|c|c|c}
Attack & Convergence & Minimum & Average \\
algorithm & time & accuracy & accuracy \\
\hline \hline
No attack & 602 & 87.40 & 88.22 \\ \hline
Attack all links & $\ge 1000$ & 56.10 & 70.14 \\ \hline
\multicolumn{4}{c}{Unlimited jamming range} \\ \hline
MCBA & $\ge 1000$ & 56.10 & 78.28 \\ \hline
NDBA & $\ge 1000$ & 56.10 & 78.28 \\ \hline
Random & $\ge 1000$ & 59.50 & 71.66 \\ \hline
\multicolumn{4}{c}{Limited jamming range} \\ \hline
MCBA & $\ge 1000$ & 56.10 & 70.40 \\ \hline
NDBA & $\ge 1000$ & 56.10 & 70.40 \\ \hline
Random & 602 & 87.40 & 88.22 
		\end{tabular}
	}
	\label{table:line}
\end{table}

Table~\ref{table:line} shows the results for the line topology.
When there is no attack, DFL can converge at time $602$ with minimum accuracy $87.40\%$. 
For the worst case that an adversary can jam all links, the models cannot converge in $1000$ rounds. The performance can be very bad for some node(s), as low as $56.10\%$ accuracy, and the average accuracy is $70.14\%$. 
For Scenario~$1$, all attacks increase the convergence time and reduce the performance.
The MCBA and the NDBA have the same performance and we find that they attack the same set of links.
The random attack is not very effective to reduce the minimum accuracy. 
For Scenario~$2$, the MCBA and the NDBA again achieve the same performance.
The random attack has the same performance as no attack because the randomly deployed jammers cannot attack any link. 

\begin{table}
	\caption{Attack performance for the star topology.}
	\centering
	{\vspace{0.2cm}
	\footnotesize
		\begin{tabular}{c|c|c|c}
Attack & Convergence & Minimum & Average \\
algorithm & time & accuracy & accuracy \\
\hline \hline
No attack & $\ge 1000$ & 69.80 & 84.30 \\ \hline
Attack all links & $\ge 1000$ & 50.40 & 80.37 \\ \hline
\multicolumn{4}{c}{Unlimited jamming range} \\ \hline
MCBA & $\ge 1000$ & 50.40 & 78.88 \\ \hline
NDBA & $\ge 1000$ & 50.40 & 78.88 \\ \hline
Random & $\ge 1000$ & 55.90 & 80.37 \\ \hline
\multicolumn{4}{c}{Limited jamming range} \\ \hline
MCBA & $\ge 1000$ & 50.90 & 64.14 \\ \hline
NDBA & $\ge 1000$ & 50.90 & 64.14 \\ \hline
Random & 270 & 69.80 & 83.88 
		\end{tabular}
	}
	\label{table:star}
\end{table}

Table~\ref{table:star} shows the results for the star topology.
DFL cannot converge in $1000$ rounds even if there is no attack. Although the average accuracy $84.30\%$ is good, the minimum accuracy is just $69.80\%$. 
For the worst case that an adversary can jam all links, the performance can be very bad for some node(s), as low as $50.40\%$ accuracy, and the average accuracy is $80.37\%$. 
For Scenario~$1$, the MCBA and the NDBA have the same performance (minimum accuracy $50.40\%$ and average accuracy $78.88\%$).
The random attack is not very effective to reduce both the minimum accuracy and the average accuracy. 
For Scenario~$2$, the MCBA and the NDBA again achieve the same performance.
The random attack converges at time $270$ with bad performance, i.e., accuracy results are still good (minimum accuracy $83.88\%$). 

\begin{table}
	\caption{Attack performance for the ring topology.}
	\centering
	{\vspace{0.2cm}
	\footnotesize
		\begin{tabular}{c|c|c|c}
Attack & Convergence & Minimum & Average \\
algorithm & time & accuracy & accuracy \\
\hline \hline
No attack & $\ge 1000$ & 78.80 & 84.11 \\ \hline
Attack all links & $\ge 1000$ & 50.40 & 63.72 \\ \hline
\multicolumn{4}{c}{Unlimited jamming range} \\ \hline
MCBA & $\ge 1000$ & 51.80 & 79.36 \\ \hline
NDBA & $\ge 1000$ & 51.80 & 79.36 \\ \hline
Random & 804 & 65.20 & 83.34 \\ \hline
\multicolumn{4}{c}{Limited jamming range} \\ \hline
MCBA & $\ge 1000$ & 51.80 & 67.21 \\ \hline
NDBA & $\ge 1000$ & 51.80 & 67.21 \\ \hline
Random & $\ge 1000$ & 58.40 & 70.98 
		\end{tabular}
	}
	\label{table:ring}
\end{table}

Table~\ref{table:ring} shows the results for the ring topology.
DFL cannot converge in $1000$ rounds even if there is no attack. The average accuracy is $84.11\%$ and the minimum accuracy is $78.80\%$. 
For the worst case that an adversary can jam all links, the performance can be very bad for some node(s), as low as $50.40\%$ accuracy, and the average accuracy is just $63.72\%$. 
For Scenario~$1$, the MCBA and the NDBA have the same performance (minimum accuracy $51.80\%$ and average accuracy $79.36\%$).
The random attack converges at time $804$ with bad performance, i.e., accuracy results are still good (average accuracy $83.35\%$). 
For Scenario~$2$, the MCBA and the NDBA again achieve the same performance.
The random attack is not very effective to reduce the average accuracy.

\begin{table}
	\caption{Attack performance for the grid topology.}
	\centering
	{\vspace{0.2cm}
	\footnotesize
		\begin{tabular}{c|c|c|c}
Attack & Convergence & Minimum & Average \\
algorithm & time & accuracy & accuracy \\
\hline \hline
No attack & 842 & 73.20 & 85.10 \\ \hline
Attack all links & $\ge 1000$ & 50.40 & 66.46 \\ \hline
\multicolumn{4}{c}{Unlimited jamming range} \\ \hline
MCBA & 761 & 51.80 & 81.75 \\ \hline
NDBA & 761 & 51.80 & 81.75 \\ \hline
Random & 558 & 68.20 & 84.90 \\ \hline
\multicolumn{4}{c}{Limited jamming range} \\ \hline
MCBA & $\ge 1000$ & 50.40 & 67.98 \\ \hline
NDBA & $\ge 1000$ & 51.80 & 69.57 \\ \hline
Random & $\ge 1000$ & 51.70 & 71.94 
		\end{tabular}
	}
	\label{table:grid}
\end{table}

Table~\ref{table:grid} shows the results for the grid topology.
DFL converges at time $842$ rounds when there is no attack. The average accuracy is $85.10\%$ and the minimum accuracy is $73.20\%$. 
For the worst case that an adversary can jam all links, DFL cannot converge in $1000$ rounds.
The performance can be very bad for some nodes(s), as low as $50.40\%$ accuracy, and the average accuracy is just $66.46\%$. 
For Scenario~$1$, the MCBA and the NDBA have the same performance (converged at time $761$ with minimum accuracy $51.80\%$ and average accuracy $81.75\%$).
The random attack converges at time $558$ but with bad performance, i.e., accuracy results are still good (average accuracy $84.90\%$). 
For Scenario~$2$, all attacks making DFL not converged in $1000$ rounds.
The MCBA is most effective (minimum accuracy $50.40\%$ and average accuracy $67.98\%$).
The NDBA has slightly higher accuracy (minimum accuracy $51.80\%$ and average accuracy $69.57\%$).
The random attack is not very effective to reduce the average accuracy.

In summary, both the MCBA and the NDBA are very effective (achieving similar results as attacking all links). Moreover, for most cases, they attack the same set of links and thus achieve the same performance. Thus, the NDBA can be adopted to simplify the attack scheme.

\section{Conclusion} 
\label{sec:conc}

In this paper, we studied a wireless signal classification problem where multiple nodes collect sensing data and train a classifier jointly in a wireless network. In a DFL scheme, each node performs federated averaging on its local model and its neighbors' models so that it can eventually collect information from all nodes via updates in multiple rounds. An adversary can launch jamming attacks on selected links to prevent model exchanges. We considered two scenarios, where a jammer may or may not limited by a jamming range. We designed both the minimum cut and the node degree based attacks, MCBA and NDBA, respectively. For various network topologies, we measure the vulnerability of DFL in wireless networks and showed that the NDBA, although simple, can achieve similar performance as the MCBA.

\end{document}